\documentclass[aps,prc,showpacs]{revtex4}

\usepackage{epsfig}
\usepackage{graphicx}

\begin{document}

\title{Precise method for the determination of the neutron electric form factor
based on a relativistic analysis of the process $d(e,e'n)p$}
\author{E. Tomasi-Gustafsson}
\affiliation{\it DAPNIA/SPhN, CEA/Saclay, 91191 Gif-sur-Yvette Cedex,
France }
\author{G.I. Gakh}
\altaffiliation{Permanent address:
\it National Science Centre "Kharkov Institute of 
Physics and Technology,\\ Akademicheskaya 1, 61108 Kharkov,
Ukraine}
\affiliation{\it DAPNIA/SPhN, CEA/Saclay, 91191 Gif-sur-Yvette Cedex,
France }

\author{A.P. Rekalo }
\altaffiliation{Permanent address:
\it National Science Centre "Kharkov Institute of 
Physics and Technology,\\ Akademicheskaya 1, 61108 Kharkov,
Ukraine}
\affiliation{\it DAPNIA/SPhN, CEA/Saclay, 91191 Gif-sur-Yvette Cedex,
France }

\author{M. P. Rekalo}
\altaffiliation{Permanent address:
\it National Science Centre "Kharkov Institute of 
Physics and Technology,\\ Akademicheskaya 1, 61108 Kharkov,
Ukraine}
\affiliation{\it DAPNIA/SPhN, CEA/Saclay, 91191 Gif-sur-Yvette Cedex,
France }
\date{\today}

\begin{abstract}
We generalize the recoil polarization method for the determination of the proton form factor to the case of the disintegration of vector polarized deuterons by longitudinally polarized electrons, $\vec d(\vec e, e'n)p$. 
We suggest to measure for this reaction, in the kinematics of quasi-elastic $en$-scattering, the ratio $R_{xz}=A_x/A_z$ of the asymmetries induced by the $x$- and $z$-components of the deuteron  vector polarization. In the framework  of the relativistic impulse approximation the ratio $R_{xz}$ is sensitive to $G_{En}$ in a wide interval of momentum transfer squared, whereas it depends weakly on the details of the $np$-interaction and on the choice of the deuteron wave function. Moreover, in the range $0.5\le Q^2\le$1.5 GeV$^2$, the ratio $R_{xz}$ shows a smooth dependence on $Q^2$, making the analysis simpler.
\end{abstract}

\maketitle
%\section{Introduction}
In the present work we are interested in the deuteron as a source of information on the neutron, in particular on its electromagnetic interaction, where the isotopic invariance is strongly violated. The most evident example is the neutron electric form factor, $G_{En}$, which shows a very different behavior when compared with $G_{Ep}$. Note in this respect the deviation of the last precise data on $G_{Ep}$ \cite{Jo99,Ga02} from the 'standard' dipole dependence on the momentum transfer squared. For a better understanding of the physics of the nucleon electromagnetic structure, more precise data about $G_{En}$ are required in a wide region of momentum transfer squared. The non-triviality of the neutron electromagnetic structure appears also in the time-like region of momentum transfer \cite{Bi90,Ba94,An94,An98}, where an unusual behavior of the cross section of the $e^+e^-$-annihilation processes has been observed:
 $$ \sigma(e^++e^-\to\bar n +n) > \sigma(e^++e^-\to\bar p+ p) \mbox{~for~} 2E=1.88\div 2.45\mbox{~GeV~}.$$

This letter is devoted to the theoretical analysis of specific polarization phenomena in the deuteron electrodisintegration, $e+d\to e+n+p$, which are  sensitive to $G_{En}$ at relatively large momentum transfer. Two kinds of  polarization experiments for $e+d\to e+n+p$ are especially interesting for this aim: the scattering of longitudinally polarized electrons by a vector polarized deuteron target, $\vec d (\vec e,en)p$, with detection of the scattered electron and  neutron \cite{PAB,ZAA}, and the scattering of longitudinally polarized electrons by an unpolarized deuteron target, with measurement of the neutron polarization, $ d (\vec e,e\vec n)p$ \cite{TE,CH,MO}. These experiments have been performed or are planned at Mainz and JLab, at momentum transfer squared  spanning from $Q^2=0.5$ to  $2$ GeV$^2$.

A correct and effective extraction of  $G_{En}$ from such data needs an  adequate theoretical interpretation of the reaction mechanism in $e+d \to e+n+p$.
First of all, the main symmetry properties of electromagnetic hadron interaction have to be taken into account, such as the conservation of the hadronic electromagnetic current (for the subprocess $\gamma^*+d\to n+p$, where $\gamma^*$ is a virtual photon), i.e. the gauge invariance of the electromagnetic interaction and the relativistic invariance. At these values of  momentum transfer relativistic effects cannot be considered as corrections, as the three-momentum of the elastically scattered nucleon is comparable with its  mass.  

The description of the final state $np$-interaction (FSI) has also to be properly taken into account. In order to decrease the model dependence of FSI , one should avoid approaches based on the concept of non-relativistic NN-potentials. Instead of NN-potentials, a model independent description of FSI in $\gamma^*+d\to n+p$ can be derived from the phases of NN-scattering, which are available from the phase-shift analysis of the huge amount of data about the NN-interaction.

The relativistically invariant impulse approximation (RIA) \cite{RGR89}, with subsequent unitarization of the corresponding multipole amplitudes \cite{U,Re94},  seems the most appropriate model for the description of  $\gamma^*+d\to n+p$. 
  
But what about the relativistic description of the deuteron structure, which can be done in terms of wave functions? Note, in this respect, that only the kinematical region for $e^-+d\to e^-+n+p$, which corresponds to quasi-elastic $e^-+n^*\to e^-+n$ scattering ($ n^*$ is a virtual neutron), is especially sensitive to neutron form factors. This region corresponds to the emission of the neutron along the three-momentum of the virtual photon, when $Q^2\simeq W^2-M^2_d$, where $M_d$ is the deuteron mass and $W$ is the effective mass of the produced $np$-system. In such conditions the virtuality of the neutron is small, therefore the argument ($p$) of the deuteron wave function (in impulse representation) is also small. In conditions of evidently nonrelativistic momentum $p$, the standard S- and D-components of the deuteron wave function (DWF), derived from the existing NN-potentials, can be safely used. Therefore,  the possible four-components of the relativistic DWF can be related with good accuracy to the nonrelativistic S- and D-components \cite{BG}.

Note that the considered kinematical regime in $\gamma^*+d\to n+p$, which is the most convenient for the determination of $G_{En}$, corresponds to nonperturbative QCD at any value of momentum transfer squared $Q^2$, so that all the prescriptions of QCD such as helicity conservation, quark counting rules, formalism of reduced deuteron form factors or reduced nuclear matrix elements cannot be applied here. Moreover,  existing experimental data, including polarization effects, concerning different processes with deuteron target: $e^-+d\to e^-+d$ \cite{Ab00}, $\gamma+d\to d+\pi^0$ \cite{Sc01}, and $\gamma+d\to n+p$ \cite{Me99} do not confirm the pQCD predictions at JLab energies.

The main aim of this paper is to show that, in the reaction 
$\vec d (\vec e,e'n)p$ (with a vector polarized deuteron target and longitudinally polarized $e^-$ beam), the information on  $G_{En}$ can be obtained in a simple and precise way from the ratio $A_x/A_z$, where $A_{x,z}$ are the corresponding asymmetries for the $x$- and $z$-components of the deuteron polarization (in the $\gamma^*+d\to n+p$ reaction plane). A similar procedure, firstly suggested in 
\cite{Re68}, has been recently realized for the processes $\vec e+p\to e+\vec p$ \cite{Jo99,Ga02}, $d(\vec e,e'\vec n p)$  \cite{CH,MO}, where the ratio of the $x$- and $z$-components of the nucleon polarization has been measured, and for the $\vec{^3{He}}(\vec e,e' n)p p$-process as well (for the ratio of the $x$-  and $z$-  components of the $\vec{^3{He}}$ polarization \cite{Me94}.

The  above mentioned ratios (in impulse approximation) for $d$ and $^3He$ targets, is essentially determined by the ratio of the electric and magnetic form factors $G_{En}/G_{Mn}$. From the experimental point of view its measurement is very convenient as many systematic errors essentially cancel and the analysis is simplified.

It is evident that in the case of polarized nuclear targets, such as $\vec d$ or $\vec{^3He}$, the ratios of target asymmetries (or final neutron polarizations) contain various nuclear effects, as well as other corrections (FSI, etc).

In the present work we update and adapt the formalism developed in \cite{RGR89}. Such formalism takes rigorously into account  important nuclear ingredients such as the Fermi motion of the nucleons in the target and the possible Wigner rotation of the vectors of the nucleon polarization. All numerical estimations are based on a relativistic treatment of the nucleon electromagnetic structure and of the FSI effects.  

%The paper is organized as follows: after a description of the formalism of %polarization phenomena in Sec. II, the results are presented in Sec. III, using different DWF and for different values of $G_{En}$, with and without FSI. In the Conclusion we summarize the results.
%\section{Formalism}

The main observable discussed here is the ratio of two $T-$even asymmetries for $\vec d(\vec e,e'n)p$, $A_x$ and $A_z$, induced by the $P_x$ and $P_z$ components of the deuteron vector polarization (in the $\gamma^*+d\to n+p$ reaction plane, with the $z$-axis along the direction of the $\gamma^*$ three-momentum $\vec q$). This ratio does not depend on the electron helicity. In conditions of quasi-elastic $en$-scattering, i.e. for $Q^2=W^2-M_d^2$ and $\vartheta^*\simeq 180^\circ$ ($\vartheta^*$ is the angle of the emitted proton with respect to $\vec q$ in the center of mass system of $\gamma^*+d\to n+p$), we show that the ratio $R_{xz}=A_x/A_z$ is sensitive to the ratio of the neutron form factors, $G_{En}/G_{Mn}$. This has been analytically proved for elastic $ep$-scattering \cite{Re68}. In the case of a deuteron target the situation is more complicated due, on one side, to the Fermi motion of bound nucleons, and from another side, to the Wigner rotation of the nucleonic spin. The final $np-$interaction plays also a role. All these aspects of the deuteron physics should be carefully taken into account for the extraction of $G_{En}$  from polarization observables in $e+d\to e+n+p$.

The asymmetries ratio can be written as follows \cite{RGR88}:
\begin{equation}
R_{xz}=\displaystyle\frac{A_x}{A_z}=\displaystyle\frac{ \sqrt{1+\epsilon}A_x^{(0)}+\sqrt{2\epsilon}A_x^{(1)}\cos\phi}
{ \sqrt{1+\epsilon}A_z^{(0)}+\sqrt{2\epsilon}A_z^{(1)}\cos\phi}, 
\label{eq:eq1}
\end{equation}
$$\epsilon^{-1}=1+\displaystyle\frac{2\vec q^2}{Q^2}\tan^2\left ( \displaystyle\frac{\theta_e}{2}\right ),$$
where $\theta_e$ is the electron scattering angle in the laboratory (Lab) system, and $\phi$ is the angle between the electron scattering plane and the reaction plane.

The specific dependence of $R_{xz}$ on $\epsilon$ and $\phi$ is a model independent result, which is based on the following properties of the hadron electromagnetic interaction:
\begin{itemize}
\item the validity of the one-photon-exchange mechanism for $e+d\to e+n+p$;
\item the conservation of the electromagnetic current for $\gamma^*+d\to n+p$ (the gauge invariance);
\item the P-invariance of hadron electrodynamics;
\item the validity of QED for the description of the $\gamma e e$-vertex.
\end{itemize}

The quantities $A_x^{(0,1)}$ and $A_z^{(0,1)}$ (the structure functions ) are real functions of three independent kinematical variables, $Q^2$, $W$, and $\vartheta^*$. Their dependence on the nucleon electromagnetic form factors can be predicted in the framework of a definite model for the process $\gamma^*+d\to n+p$. 

Note that $A_x^{(0)}$ and $A_z^{(0)}$ are determined by quadratic combinations of the transversal components of the electromagnetic current for $\gamma^*+d\to n+p$, whereas $A_x^{(1)}$ and $A_z^{(1)}$ are driven by the interference of the longitudinal and the transversal components of this current. Moreover,  $A_x^{(0)}$ and $A_z^{(1)}$ vanish at $\vartheta^*=0^{\circ}$ and $180^{\circ}$, due to the helicity conservation for  collinear kinematics in  $\gamma^*+d\to n+p$. These results are also model independent. 

However, considering the one-nucleon contributions of the electromagnetic current for $\gamma^*+d\to n+p$, one can expect that its longitudinal component is determined by $G_{En}$ and the transversal one by $G_{Mn}$, at least in the Breit system. Therefore, the contributions $A_{x,z}^{(1)}$, in Eq. (\ref{eq:eq1}), should contain terms proportional to the product $G_{En}G_{Mn}$.

The presence of two contributions in both asymmetries,  $A_x$ and $A_z$, results from the specific character of the deuteron dynamics for $\gamma^*+d\to n+p$ and their relative role is determined by the corresponding model. 
\begin{figure}
\begin{center}
\includegraphics[width=7cm]{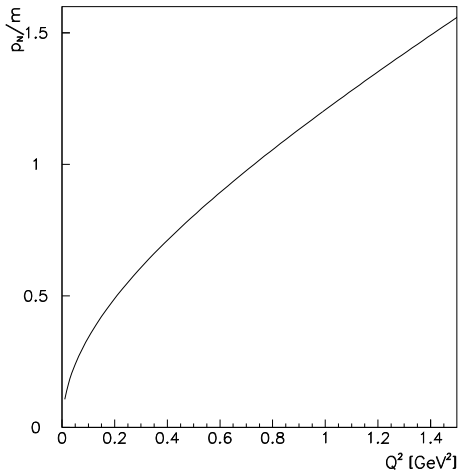}
\caption{\label{fig:fig1} $Q^2$-dependence of the nucleon momentum in $d(e,e'n)p$, in quasi-elastic kinematics (Lab system).
}
\end{center}
\end{figure}

\begin{figure}
\begin{center}
\includegraphics[width=8cm]{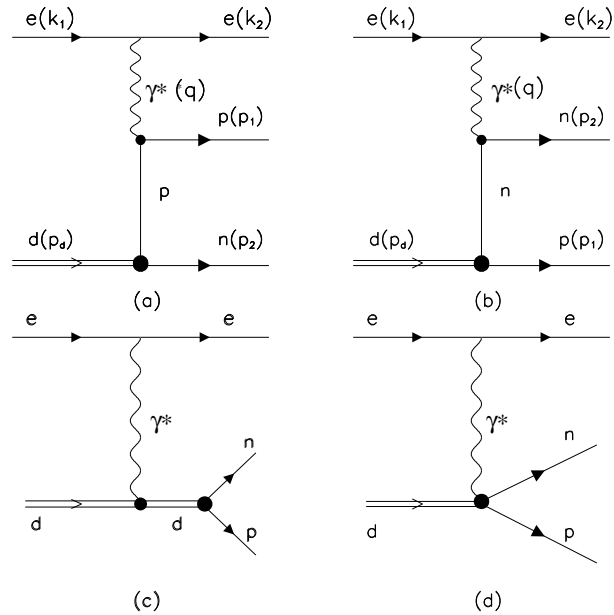}
\caption{\label{fig:fig2} Feynman diagrams describing RIA for the process $e+d\to e+n+p$.
}
\end{center}
\end{figure}

A realistic model has to be relativistic, due to the fast rising of the emitted nucleon momentum, $p_N$, with increasing $Q^2$, which can be written in the Lab  system as:
$$\displaystyle\frac{p_N}{m}=2\sqrt{\tau(1+\tau)},~
\tau=\displaystyle\frac{Q^2}{4m^2},$$
in quasi-elastic kinematics (see Fig. \ref{fig:fig1}), where $m$ is the nucleon mass. Therefore, the corresponding impulse approximation can be realized by means of the Feynman technique, in which all particles 
are treated relativistically, namely: the nucleons are described by the 
four-components Dirac spinors, and deuteron - by the polarization 4-vector 
$U_{\mu}$. The diagrams illustrated in  Fig. \ref{fig:fig2} determine the amplitude 
in RIA.  Comparing with a  nonrelativistic approach, the diagrams (a) and (b) represent the relativized description of the one--nucleon--exchange 
mechanism (which are equally important in both models). The deuteron--exchange  (c) as well as the contact diagram  (d) insure 
the electromagnetic current conservation in $\gamma ^*+d\rightarrow n+p$. The contact diagram 
can be related to the contribution of the meson--exchange currents. Of course, 
this diagram is not comprehensive of all variety of these currents,
but for its structure and origin it falls into this class. The deuteron diagram 
can be also related to the $np$-interaction  in definite states, with the quantum numbers of the deuteron.
The deuteron structure (which, in the nonrelativistic approach, corresponds to DWF) is described here by the relativistic form 
factors of the $dnp-$ vertex with one virtual nucleon \cite{BG}. In order to 
calculate the dependence of these form factors on the nucleon virtuality we 
use the relation between them and nonrelativistic DWF.

The RIA amplitude for $\gamma^*+d\rightarrow n+p$ reaction can be 
unambiguously calculated for any values of the kinematic variables $Q^2$, $W$ and $\vartheta^*$.

In the case of the nonrelativistic description of the deuteron
electrodisintegration the situation is somewhat different. The standard
nuclear approach to the investigation of the electron scattering by deuterons
and other nuclei requires the knowledge of the operator of the electron--nucleon
interaction. This operator cannot contain the antinucleon contribution
which is inevitably present in the covariant description of the $eN$-
scattering (with one virtual nucleon). The method based on the Foldy-Wouthoysen transformation is usually
applied. It allows to obtain this operator in the form of the expansion over the
powers of $\sqrt{Q^2}/m$. Naturally, this method is not valid at $Q^2\ge m^2$, but such values of momentum transfer are currently accessible. Therefore, a relativistic description of the 
$e^-+d\rightarrow e^-+n+p$ reaction is more appropriate. FSI have also to be taken properly into account. Instead of the nonrelativistic concept of $NN$-potential (in any form), which is typically assumed for nuclear processes, even at large $Q^2$, it is more straightforward to use the phases of $NN$-scattering, which are available now up to E=0.8 GeV for $np$-scattering. Generalizing the Fermi-Watson approach, we will use the following unitarization procedure: 
\begin{equation}
f_{[A]}^{(0)}(Q^2, W)\longrightarrow f_{[A]}^{(0)}(Q^2, W)
exp(i\delta _{[A]}(W)),
\label{eq:eq2}
\end{equation}
where $f_{[A]}^{(0)}(Q^2, W)$ is an RIA multipole amplitude for the process 
$\gamma ^* +d\rightarrow n+p$ leading to the production of the $np$ system in a 
state with quantum numbers $[A]=J$ (total angular momentum), $L$ (orbital 
angular momentum), $S$ (total spin), and $I$ (total isotopic spin), 
while $\delta _{[A]}(W)$ is the
NN phase shift in the state with quantum numbers $[A].$ This expression is
strictly valid in the energy range $2m\leq W\leq (2m+m_{\pi})$ at any value
of $Q^2$ in the region of space-like momentum transfers.

The substitution given by Eq. (\ref{eq:eq2}) is performed only for those multipole
amplitudes that describe the production of the $np$ system with nonzero phase
shift $\delta _{[A]}(W)$. This means that, at each energy $W$, there exist the
maximum angular--momentum value $J=J_m(W)$ (more precisely, maximum value
of $L$ for the $np$ system) that limits the number of those 
multipole amplitudes that are modified by the unitarization. A similar "cut" on $L$ is imposed by a 
finite (and small) range of $NN$ interaction. Even despite this constrain, 
it is necessary to consider and modify a rather large number of multipole amplitudes. At $J=0 (J=1)$, there are 3 (14) independent transitions, so that it is
necessary to modify $18(J-1)+14+3=18J-1$ independent multipole amplitudes. For $J_m\geq 6$ their number exceeds one hundred.
The unitarization procedure is carried out in a relativistic
approach, without any restrictions on $Q^2$. 

The nucleon electromagnetic current is described in terms of 
the Dirac ($F_1$) and Pauli ($F_2$) form factors. For the $np$ phase shifts, we took 
the results from \cite{Bys}, where the phases have been found up to $W\simeq $ 2.24 GeV.
\vspace{0.5cm}
\begin{table}[h]
\begin{tabular}{|c||c||c||c||c|}
\hline
$Q^2 [GeV^2]$ & E [GeV] & E' [GeV] & $\theta _e$ [deg] & 
$\varepsilon$ \\
\hline
\it 0.5 & 2.721 & 2.460 & 15.8 & 0.958\\
\hline
\it 1.0 & 4.232 & 3.698 & 14.5 & 0.960\\
\hline
\it 1.5 & 4.232 & 3.376 & 19.26 & 0.923\\
\hline
\end{tabular}
\caption{Table1: Kinematical parameters for the JLab experiment \protect\cite{Day}.}
\end{table}
%%%%%%%%%%%%%%%%%%%%%%%%%%%%%%%
%\section{Results}
%%%%%%%%%%%%%%%%%%%%%%%%%%%%%%%

We calculated the ratio $R_{xz}$ for the electron kinematics reported in Table I, which correspond (at $Q^2$=0.5 and 1 GeV$^2$ to the kinematical conditions of the experiment E93-026 recently performed at JLab \cite{Day}. Note that $\epsilon\simeq 1$ for the quasi-elastic regime.

We consider the $\vartheta^*$-dependence of $R_{xz}$ in the interval $\vartheta^*=180\pm 20^0$, and use the relation $W^2=Q^2+M_d^2$ to determine the variable $W$ in the quasi-elastic $en$-region. The nucleon electromagnetic form factors are taken as follows:
$$G_{Ep}(Q^2)=G_{Mp}(Q^2)/\mu_p=G_{Mn}(Q^2)/\mu_n=G_D=(1+Q^2/0.71\mbox{~GeV}^2)^{-2},$$
$$
G_{En}=G^G=-\tau\mu_nG_D/(1+5.6\tau),$$
where $\mu_p(\mu_n)$ is the magnetic moment of the proton (neutron).
\begin{figure}
\begin{center}
\includegraphics[width=15cm]{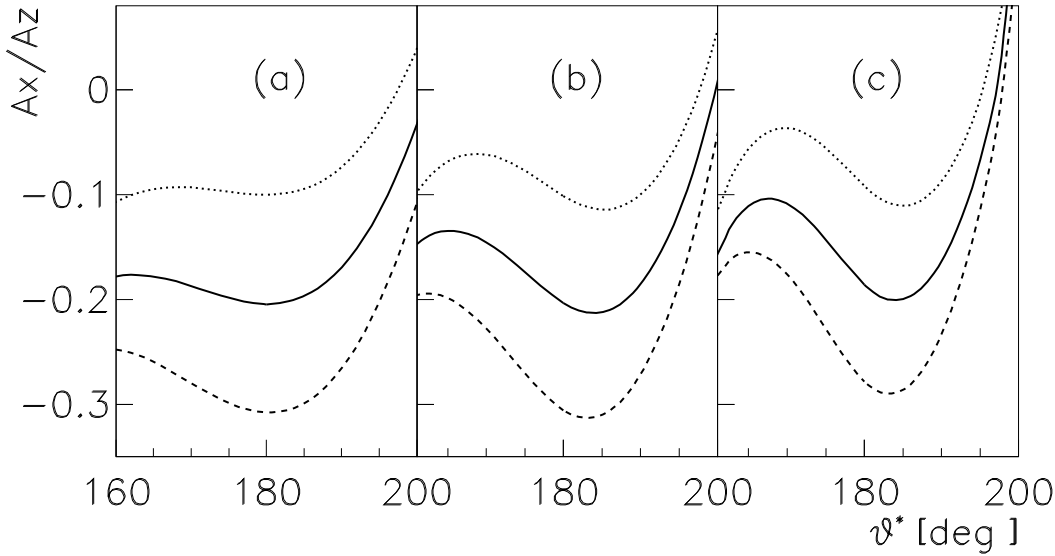}
\caption{\label{fig:fig3} $\vartheta^*$-dependence of the ratio $R_{xz}=A_x/A_z$ for different values of the neutron electric form factor: Galster parametrization (solid line), Galster parametrization scaled by a factor 0.5 (dashed line), Galster parametrization scaled by a factor 1.5 (dotted line)
 at $Q^2$=0.5 GeV$^2$ (a); $Q^2$=1 GeV$^2$ (b); $Q^2$=1.5 GeV$^2$ (c); $\vartheta^*<180^{\circ}$ ($>180^{\circ}$) correspond to $\phi=0$ ($180^{\circ}$).}
\end{center}
\end{figure}
In Fig. \ref{fig:fig3} the results for $R_{xz}$ are shown at $Q^2$ =0.5, 1, and  1.5 GeV$^2$. The calculation corresponding to the Paris DWF \cite{La80} and to the Galster parametrization for $G_{En}$ \cite{Ga71} is shown as solid line. The calculations corresponding to $G_{En}=0.5~ G^G$ (dashed line) and  $G_{En}=1.5 ~G^G$ (dotted line) are also reported, in order to show the large sensitivity of the ratio $R_{xz}$ to $G_{En}$, in the considered $Q^2$ and $\vartheta^*$ range. Only at $Q^2=0.5$ GeV$^2$ a small effect of FSI and DWF appears at $\vartheta^*\simeq 160^\circ$, where the cross section is smaller than in the quasi-elastic region. At larger $Q^2$, the effects of FSI and DWF are negligible, making simpler the extraction of $G_{En}$ from  experimental data on $R_{xz}$ .

\begin{figure}
\begin{center}
\includegraphics[width=15cm]{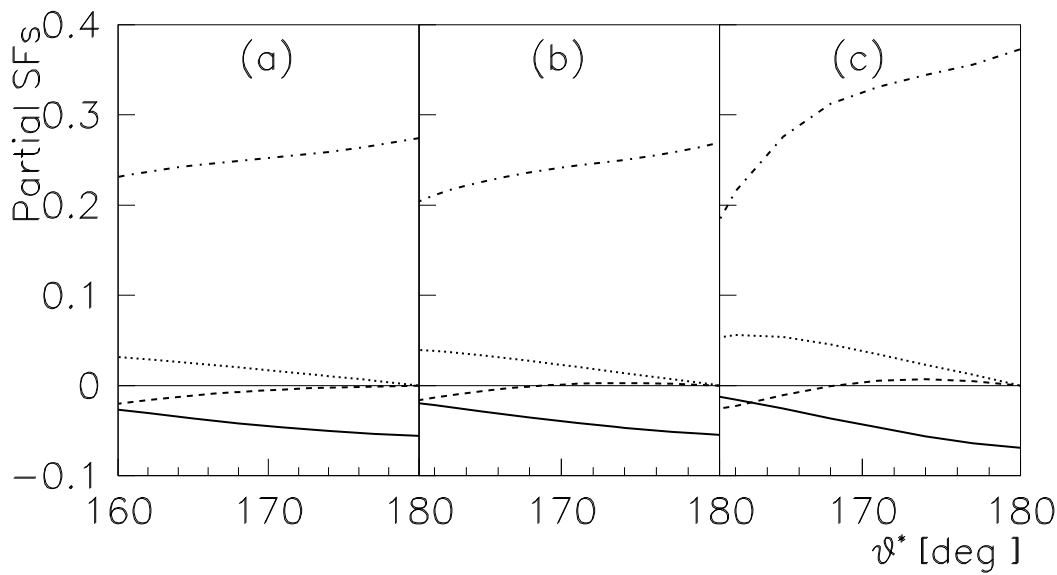}
\caption{\label{fig:fig4} $\vartheta^*$-dependence of the partial structure functions $A_x^{(0)}$  (dashed line), $A_x^{(1)}$ (solid line), $A_z^{(0)}$ (dash-dotted line), and $A_z^{(1)}$ (dotted line) 
for $Q^2$=0.5 GeV$^2$ (a); $Q^2$=1 GeV$^2$ (b); $Q^2$=1.5 GeV$^2$ (c)}.
\end{center}
\end{figure}

The relative role of the two possible contributions to $A_x$ and $A_z$ (see Eq. (\ref{eq:eq1})) is shown in Fig. \ref{fig:fig4}. For $\vartheta^*\ne 180^\circ$ these contributions are comparable for $A_x$, being negative for $\vartheta^* < 180^\circ$. For $A_z$, the transversal contribution $A_z^{(0)}$ , related to $G_{Mn}^2$, is essentially larger in comparison to $A_z^{(1)}$. 

%\section{Conclusions}

In conclusion, the suggested method for the determination of $G_{En}$, at relatively large $Q^2$, from the ratio $R_{xz}=A_x/A_z$ of the T-even asymmetries in $\vec d(\vec e,e'n)p$, measured in the kinematical conditions of quasi-elastic $en$- scattering, seems promising and may be comparable in accuracy with the measurements of $G_{Ep}$ through the recoil polarization method. The nuclear effects, such as the Fermi motion and the Wigner rotation of the nucleon polarization, can be taken rigorously into account through a formalism based on RIA. The final $np$--interaction and the choice of the deuteron wave function do not influence the suggested procedure.

This method would be especially useful when the direction of the target polarization can be done rapidly, allowing a simultaneous measurement of the two asymmetries. It is the case of internal gas targets \cite{Bates}.

For external targets, the procedure of rotating the target polarization may involve complicated manipulations of the magnets and the experimental supplies, making the measurements with the two polarization states of the target sequential in time. Although, in this case some parameters may change, the ratio of the asymmetries seems more reliable for the extraction of $G_{En}$, as it is less sensitive to the models than the individual asymmetries.

%\section{Acknowledgments}
Three of us, (G.I.G., A.P.R., and M.P.R.) acknowledge the kind hospitality of Saclay, where this work was done. Thanks are due to D. Day  for useful discussions and remarks.
{}

\vspace{0.5cm}
\newpage

\newpage

\end{document}